# RESULTS FROM LATTICE QCD FOR BARYONS CONTAINING A HEAVY QUARK

UKQCD Collaboration

Presented by P. Ueberholz

*Department of Physics and Astronomy, The University of Edinburgh*
*The Kings's Buildings, Edinburgh, EH9 3JZ, UK*

We present the calculation of the spectrum of baryons containing one heavy quark. Heavy baryon and meson mass splittings are computed and compared with experiment. We give preliminary results for the form factor $G_1$ for the semileptonic decay $\Lambda_b \to \Lambda_c l \bar{\nu}$ and investigate its flavour symmetry.

## 1 Introduction

Hadrons containing a heavy quark and two light quarks can be studied succesfully via lattice QCD calculations, which provide non perturbative, model-independent results. Recently there have been two attempts to compute the mass of the $\Lambda$-baryon as well as other baryons containing heavy quarks on the lattice[1,2]. A lattice analyses of the hadronic matrix elements of the semi-leptonic decay $\Lambda_b \to \Lambda_c$ should enable an additional independent measurement of the element $V_{cb}$ of the CKM matrix. A similar study was already carried out in the framework of Heavy Quark Effective Theory (HQET), (for a review see[3]). We present the theoretical framework of both spectroscopy and weak matrix elements in section 2, and report our results in section 3.

## 2 Lattice Operators for Heavy Baryons

There are eight baryons containing one heavy and two light quarks (see table 1). Their spectrum can be computed on the lattice by using the following operators:

- $\Lambda_h, \Xi_h \to \mathcal{O}_5 = \epsilon_{abc}(l^a \mathcal{C} \gamma_5 l'^b) h^c$
- $\Sigma_h, \Sigma_h^*, \Omega_h, \Omega_h^* \to \mathcal{O}_\mu = \epsilon_{abc}(l^a \mathcal{C} \gamma_\mu l^b) h^c$
- $\Xi_h^*, \Xi_h', \to \mathcal{O}_\mu' = \epsilon_{abc}(l^a \mathcal{C} \gamma_\mu l'^b) h^c$

where $\mathcal{C}$ is the charge conjugation matrix, $l, l'$ are light quark fields and $h$ is the heavy quark field.

The operator $\mathcal{O}_5$ corresponds to $s_l^{\pi_l} = 0^+$ spin-parity for the light degrees of freedom and a total spin-parity for the baryon $J^P = \frac{1}{2}^+$. We can extract the masses of the corresponding baryons for large time separations from the correlation function

$$G_5(\vec{p},t) = \sum_{\vec{x}} e^{-i\vec{p}\cdot\vec{x}} \left\langle \mathcal{O}_5(\vec{x},t)\overline{\mathcal{O}}_5(\vec{0},0) \right\rangle$$
$$\to \frac{Z^2(|\vec{p}|)}{2E} e^{-Et}(\not{p}+M). \quad (1)$$

The operator $\mathcal{O}_\mu^{(\prime)}$ can create both spin-$\frac{3}{2}$ and spin-$\frac{1}{2}$ particles. By defining[4] the projectors $P^{3/2}$ and $P^{1/2}$ one can separate the $\frac{3}{2}$ and $\frac{1}{2}$ components of the correlation functions of the operator $\mathcal{O}_\mu$

$$G_{\mu\nu}(\vec{p},t) = \sum_{\vec{x}} e^{-i\vec{p}\cdot\vec{x}} <\mathcal{O}_\mu(\vec{x},t)\overline{\mathcal{O}}_\nu(\vec{0},0)>$$
$$\to \frac{Z_{\frac{3}{2}}^2}{2E_{\frac{3}{2}}} e^{-E_{\frac{3}{2}}t}(\not{p}+M_{\frac{3}{2}})P_{\mu\nu}^{\frac{3}{2}} + \frac{Z_{\frac{1}{2}}^2}{2E_{\frac{1}{2}}} e^{-E_{\frac{1}{2}}t}(\not{p}+M_{\frac{1}{2}})P_{\mu\nu}^{\frac{1}{2}}.$$

The process $\Lambda_b \to \Lambda_c l \bar{\nu}_l$ is mediated by the weak current $(J_\mu(y))^{h' \to h} = \bar{h}(y)\gamma_\mu(1-\gamma_5)h'(y)$, whose matrix elements between baryonic states are parametrized by six form factors, $F_i$ for the vector and $G_i$ for the axial current respectively, as follows

$$\langle \Lambda_h^s(\vec{p}) | (J_\mu(y))^{h' \to h} | \Lambda_{h'}^{s'}(\vec{p}') \rangle$$
$$= \bar{u}_h^{(s)}(\vec{p}) \mathcal{F}_\mu^{h,h'}(p',p) u_{h'}^{(s')}(\vec{p}') \quad (2)$$

with

$$\mathcal{F}_\mu^{h,h'}(p',p) = \left(F_1(\omega)\gamma_\mu + F_2(\omega)v_\mu' + F_3(\omega)v_\mu\right)$$
$$- \left(G_1(\omega)\gamma_\mu + G_2(\omega)v_\mu' + G_3(\omega)v_\mu\right)\gamma_5 \quad (3)$$

where $u_{h^{(\prime)}}^{(s^{(\prime)})}(\vec{p}^{(\prime)})$ are the spinors of the heavy baryons, $s^{(\prime)}$ are helicity indices, $v_\mu^{(\prime)} = p_\mu^{(\prime)}/M_{\Lambda_{h^{(\prime)}}}$ and $\omega = v \cdot v'$. HQET imposes strong restrictions on the form factors. They are related to one universal mass and renormalization scheme independent function $\zeta$ of the velocity transfer, $\omega$, which we call the baryonic Isgur-Wise function:

$$F_i(\omega) = (\alpha_i + \beta_i(\omega) + \gamma_i(\omega))\zeta(\omega) \quad (4)$$
$$G_i(\omega) = \left(\alpha_i^5 + \beta_i^5(\omega) + \gamma_i^5(\omega)\right)\zeta(\omega) \quad (5)$$
$$\alpha_1 = \alpha_1^5 = 1 \qquad \alpha_{2,3} = \alpha_{2,3}^5 = 0.$$

$\beta_i(\omega), \beta_i^5(\omega)$ represent radiative corrections and $\gamma_i(\omega), \gamma_i^5(\omega)$ correspond to corrections proportional to the in-verse powers of the heavy quark masses. The baryonic Isgur-Wise function is normalized at zero recoil: $\zeta(\omega = 1) = 1$.

The quantities $\sum_i F_i(\omega)$ and $G_1(\omega)$ have no $1/m_Q$ corrections at zero recoil. Therefore it should be possible to extract an accurate value of $V_{cb}$ from the measurement of semi-leptonic $\Lambda_b$ decays near zero recoil, where the decay rate is governed only by the form factor $G_1$

$$\lim_{\omega \to 1} \frac{1}{\sqrt{\omega^2 - 1}} \frac{d\Gamma(\Lambda_b \to \Lambda_c l \bar{\nu}_l)}{d\omega}$$
$$= \frac{G_F^2 |V_{cb}|^2}{4\pi^3} m_{\Lambda_c}^3 (m_{\Lambda_b} - m_{\Lambda_c})^2 |G_1(1)|^2. \quad (6)$$

In order to study this matrix element on the lattice, we consider the following three-point correlators:

$$(C_{\alpha\beta}(t_x, t_y))_\mu^{h' \to h} \quad (7)$$
$$= \sum_{\vec{x}, \vec{y}} e^{-i\vec{p}\cdot\vec{x}} e^{-i\vec{q}\cdot\vec{y}} \langle \mathcal{O}_5^{\alpha h}(x) (J_\mu(y))^{h' \to h} \bar{\mathcal{O}}_5^{\beta h'}(0) \rangle.$$

Providing both $t_y, t_x - t_y$ are large, the ground state dominates and the correlator can be expressed in terms of the lattice form factors $F_{i,lat}(\omega)$ and $G_{i,lat}(\omega)$. They are related to the continuum form factors by multiplicative renormalization constants, up to discretization errors.

$$(C_{\alpha\beta}(t_x, t_y))_\mu^{h' \to h} = \frac{Z_5^h(\vec{p})}{2E^h(\vec{p})} \frac{Z_5^{h'}(\vec{p}+\vec{q})}{2E^{h'}(\vec{p}+\vec{q})}$$
$$\times e^{-E^h(\vec{p})(t_x - t_y)} e^{-E^{h'}(\vec{p}+\vec{q})t_y}$$
$$\times \left( (\slashed{p}_h + M_{\Lambda_h}) \mathcal{F}_{\mu lat.}^{h,h'} (\slashed{p}_{h'}' + M_{\Lambda_{h'}}) \right)_{\alpha\beta} \quad (8)$$

where $\mathcal{F}_{\mu lat.}^{h,h'}$ is the lattice analogue of the function introduced in eq. (3), and $p' = p + q$. The wave-function factors, $Z_5^{h,h'}$, and energies can be obtained from the analysis of the appropriate two-point functions.

## 3 Details of the Simulation

We used the $O(a)-$improved fermion action, introduced by Sheikholeslami and Wohlert (SW)[5]. It has been shown[6] that, when the SW action is used in conjunction with a 'rotation' of the quark fields, all the $O((g_0)^2 a \log a)$ effects are removed from hadronic matrix elements. In order to isolate the ground state in correlation functions effectively, we used extended (smeared) interpolating operators[7].

The analysis is based on 60 configurations at $\beta = 6.2$ of a $24^3 \times 48$ lattice. We computed the correlation functions for four values of the heavy-quark mass, $\kappa_h = 0.133, 0.129, 0.125, 0.121$, and two light-quark masses $\kappa_l = 0.14226, 0.14144$. With these two light-quark masses we computed two degenerate and one non-degenerate light combinations. $\kappa_{crit} = 0.14315(2)$ corresponds to vanishing quark mass, $\kappa_s = 0.1419(1)$ to the strange and $\kappa_c = 0.129$ approximately to the charm quark mass.

## 4 Baryon masses

We computed the correlators given in eq.(1) and eq.(2) and modelled the extrapolation both in the light-quark and in the heavy-quark masses[8] by linear functions. In order to convert our values for baryon masses into physical units we need an estimate of the inverse lattice spacing. In this study we take

$$a^{-1} = 2.9 \pm 0.2 \text{ GeV} \quad (9)$$

The error in eq.(9) is large enough to encompass all our estimates for $a^{-1}$ from quantities such as $m_\rho, f_\pi, r_0, m_N$ and the string tension. The results for the heavy baryon masses are listed in table 1.

Table 1: Lattice determination of the heavy baryon masses, together with the experimental data, where it exists.

| Baryon $h = c, b$ | $J^P$ | Mass [GeV] | Quark Content | Lattice [GeV] |
|---|---|---|---|---|
| $\Lambda_h$ | $\frac{1}{2}^+$ | 2.285(1) | $(ud)c$ | 2.28 $^{+4}_{-4}$ |
| | | 5.64(5) | $(ud)b$ | 5.59 $^{+9}_{-10}$ |
| $\Sigma_h$ | $\frac{1}{2}^+$ | 2.453(1) | $(uu)c$ | 2.45 $^{+6}_{-4}$ |
| | | 5.81(6) | $(uu)b$ | 5.69 $^{+10}_{-10}$ |
| $\Sigma_h^*$ | $\frac{3}{2}^+$ | 2.53(1) | $(uu)c$ | 2.43 $^{+5}_{-4}$ |
| | | 5.87(6) | $(uu)b$ | 5.68 $^{+9}_{-10}$ |
| $\Xi_h$ | $\frac{1}{2}^+$ | 2.468(4) | $(us)c$ | 2.41 $^{+3}_{-3}$ |
| | | | $(us)b$ | 5.69 $^{+7}_{-10}$ |
| $\Xi_h'$ | $\frac{1}{2}^+$ | | $(us)c$ | 2.58 $^{+5}_{-4}$ |
| | | | $(us)b$ | 5.85 $^{+10}_{-12}$ |
| $\Xi_h^*$ | $\frac{3}{2}^+$ | 2.643(2) | $(us)c$ | 2.55 $^{+5}_{-4}$ |
| | | | $(us)b$ | 5.82 $^{+8}_{-11}$ |
| $\Omega_h$ | $\frac{1}{2}^+$ | 2.70(2) | $(ss)c$ | 2.68 $^{+4}_{-4}$ |
| | | | $(ss)b$ | 5.91 $^{+7}_{-8}$ |
| $\Omega_h^*$ | $\frac{3}{2}^+$ | | $(ss)c$ | 2.64 $^{+4}_{-3}$ |
| | | | $(ss)b$ | 5.89 $^{+6}_{-8}$ |

We measured the $m_\Lambda - m_P$ splitting taking the difference of the fitted masses, whereas $m_\Sigma - m_\Lambda$ has been fitted from the ratio of the corresponding correlators. The extrapolations of the mass splittings, $\Delta$, have been modelled by a linear function of the inverse heavy pseudoscalar mass, according to the HQET predictions:

$$\Delta = A + \frac{B}{m_P(\kappa_h)}. \quad (10)$$

Here we present results for the dimensionless quantities

$$R(\Lambda) = \frac{m_\Lambda - m_P}{m_\Lambda + m_P} \qquad R(\Sigma) = \frac{m_\Sigma - m_\Lambda}{m_\Sigma + m_\Lambda}. \qquad (11)$$

We get:

|  | Exp. | Lattice |
|---|---|---|
| R($\Lambda_c$) | 0.100(3) | 0.099 $^{+9}_{-7}$ |
| R($\Lambda_b$) | 0.033(5) | 0.033 $^{+5}_{-4}$ |
| R($\Sigma_c$) | 0.035(1) | 0.038 $^{+9}_{-9}$ |
| R($\Sigma_b$) | 0.016(2) | 0.017 $^{+5}_{-7}$ |

The splittings of the spin doublets $(\Sigma^*, \Sigma), (\Xi^*, \Xi)$ and $(\Omega^*, \Omega)$ are negative although with a large statistical error that encompasses zero in most cases. Furthermore the spin splitting may be subject to large systematic effects.

## 5 Form Factor $G_1$

The form factors are computed on the lattice by analysing the quantity

$$R_{\alpha\beta}(t_x, t_y))_\mu^{h' \to h} = \frac{(C_{\alpha\beta}(t_x, t_y))_\mu^{h' \to h}}{G_5(t_x - t_y)^h G_5(t_y)^{h'}} \qquad (12)$$

which tends to a constant proportional to the lattice matrix element for $0 \ll t_y \ll t_x \ll L$. We constrain the remaining factors from the 2-point function and hence determine $G_{1,lat}$.

In the case of degenerate transitions, in which the heavy quarks are of equal mass, we reduce discretization errors as well as remove the dependence on the matching coefficient between the lattice and continuum form factor by dividing $G_{1,lat.}(\omega)$ by the measured $G_{1,lat.}(1)$. The denominator is obtained from the transition $(0,0,0) \to (0,0,0)$; the error bars at zero recoil arise from the measurement of the transition $(1,0,0) \to (1,0,0)$.

We present in fig. 1 preliminary results for $G_1(\omega)$ normalized to $G_1(1) = 1$ for the degenerate transition at our four values of $\kappa_h$ with both light quarks corresponding to $\kappa_l = 0.14144$.

Radiative corrections are not included in this preliminary analysis so some caution has to be exercised before calling this the "Isgur-Wise" function. The power corrections ($\gamma_i$) are non-perturbative and are not yet determined in a model-independent way. Since $G_1$ has no $O(1/m_Q)$ correction at zero recoil we might expect that these corrections are small. The points appear to lie on a universal curve suggesting that such mass-dependent corrections are indeed small

## 6 Conclusions

The full spectrum of the eight heavy baryons containing a single heavy quark ($c$ or $b$) has been computed.

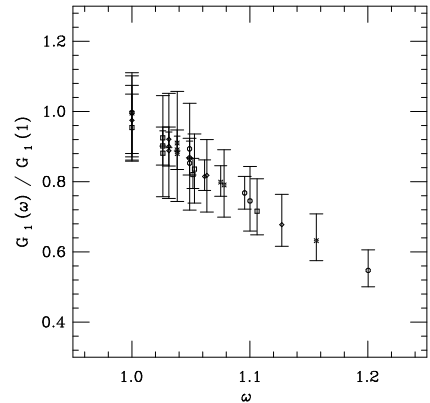

Figure 1: The form factor $G_{1,lat.}(\omega)/G_{1,lat.}(1)$ for degenerate transitions. Different symbols correspond to different $\kappa_h$-values.

The extrapolation to the chiral limit as well as in the heavy quark mass looks very reliable and the agreement between our estimates and the physical values is good. The estimates for $M_\Lambda - M_P$ and $M_\Sigma - M_\Lambda$ agree well with the experimental data, but the measurement of the spin splitting $M_{\Sigma^*} - M_\Sigma$ is inconclusive. A more extensive discussion of the results can be found in[9].

We have computed the form factor $G_{1,lat}$. We obtain a good signal for the form factor and are able to extract its $\omega$ dependence. This study confirms that the dependence of $G_1$ on the heavy quark mass is very weak, and undetectable within the precision of our data.

## Acknowledgments

This work was supported by the Particle Physics and Astronomy Research Council. PU acknowledges the support of the HCM project of the EC, contract ERBCH-BICT930877.